\documentstyle[epsfig,12pt]{article}

\def\thefootnote{\fnsymbol{footnote}}
\newlength{\minitwocolumn}
\begin{document}

\newcommand{\dfrac}[2]{\displaystyle{\frac{#1}{#2}}}

\def\thefootnote{\fnsymbol{footnote}}

\hspace*{10cm} {\bf AMU-97-01}\\[-.3in]

\hspace*{10cm} {\bf US-97-03}\\[-.3in]

\hspace*{10cm} {\bf June 1997}\\[.4in]


\begin{center}

{\large\bf Updated Running Quark Mass Values}
\footnote{
Talk presented by H.~Fusaoka at the  1997 Shizuoka Workshop 
on Masses and Mixings of Quarks and Leptons,  
Shizuoka, Japan, March 1997.  To appear in the Proceedings.}
\\[.3in]

{\bf Hideo Fusaoka}\footnote{
E-mail: fusaoka@amugw.aichi-med-u.ac.jp} \\

Department of Physics, Aichi Medical University \\ 
Nagakute, Aichi 480-11, Japan \\[.1in]

and \\[.1in]

{\bf Yoshio Koide\footnote{
E-mail address: koide@u-shizuoka-ken.ac.jp}} \\[.1in]

Department of Physics, University of Shizuoka \\[.1in] 
52-1 Yada, Shizuoka 422, Japan \\[.5in]

{\large\bf Abstract}\\[.1in]
\end{center}

\begin{quotation}
The running quark masses $m_q(\mu)$ at 
various energy scales $\mu$ ($\mu = 1 {\rm GeV}, \ \mu = m_q, \ \mu = m_Z$ 
and so on) are evaluated by using the mass renormalization equations 
systematically. 
Also, those at energy scales $\mu$ higher than $\mu = m_Z$ (from 
$\mu = 10^3$ GeV to $\mu = 10^{16}$ GeV) are evaluated by using the 
evolution equations of the Yukawa coupling constants for the standard model
with a single Higgs boson.
\end{quotation}

\newpage

\noindent {\bf 1. Introduction}

\vglue.1in

It is very important to know reliable values of quark masses $m_q$ not 
only for hadron physicists  but also for quark-lepton physicists. 
For such a purpose, a review article by Gasser and 
Leutwyler [1] has offered useful information on the running quark masses 
$m_q(\mu)$. 
However, since Gasser and Leutwyler's review article [1] in 1982, 
there have been some developments, for example,
higher order calculation of perturbative QCD [2, 3],
matching condition at quark threshold [4],
and discovery of the top quark [5].

In this talk we report the running quark masses $m_q(\mu)$ at 
various energy scales $\mu$ ($\mu = 1 {\rm GeV}, \ \mu = m_q, \ \mu = m_Z$ 
and so on) which are evaluated by using the mass renormalization 
equations systematically. 
The calculation was done by taking the matching condition at the 
quark flavor threshold into account.
Also, those at energy scales $\mu$ higher than $\mu = m_Z$ (from 
$\mu = 10^3$ GeV to $\mu = 10^{16}$ GeV) are evaluated by using the 
evolution equations of the Yukawa coupling constants for the standard model
with a single Higgs boson.

In the next section, we review  values of light quark masses 
$m_u(\mu)$, $m_d(\mu)$ and $m_s(\mu)$ at $\mu=1$ GeV. 
In Sec.3, we review pole mass values of heavy quark masses $M_c^{pole}$, 
$M_b^{pole}$ and $M_t^{pole}$. In order to estimate 
$m_q(\mu)$ at any $\mu$, we must know the values of the QCD 
parameters $\Lambda^{(n)}_{\overline{\rm MS}}$ ($n=3,4,5,6$).
In Sec.4,  the values of $\alpha_s(\mu)$ and 
$\Lambda_{\overline{\rm MS}}^{(n)}$ are reviewed.
In Sec.5, running quark masses $m_q(\mu)$ are evaluated for various energy 
scales $\mu$, $e.g.$, 
$\mu = 1$ GeV, $\mu = m_q$, $\mu = M_q^{pole}$, $\mu = m_Z$, 
$\mu = \Lambda_W$, and so on, where $M_q^{pole}$ is a ^^ ^^ pole" mass of the 
quark, and $\Lambda_W$ is a symmetry breaking energy scale of 
the electroweak gauge symmetry SU(2)$_L$ $\times$ U(1)$_Y$, i.e.
$
\Lambda_W \equiv \langle \phi^0 \rangle = (\sqrt{2}G_F)^{-\frac{1}{2}}/ 
\sqrt{2} = 174.1 {\rm GeV} 
$.
In Sec.6, the reliability of the perturbative calculation 
below $\mu\sim 1$ GeV is discussed.
In Sec.7, evolution of the Yukawa coupling constants of the standard model
with a single Higgs boson is estimated 
for energy scales higher than $\mu=\Lambda_W$. 
Finally, Sec.8 is devoted to summary and discussion. 

\vspace*{.2in}

\noindent {\bf 2. Review: light quark masses at $\mu=1$ {\rm GeV}}

\vglue.1in

Since Gasser and Leutwyler [1] have obtained the light quark masses 
$m_u(\mu)$, $m_d(\mu)$ and $m_s(\mu)$ at $\mu=1$ GeV, 
various values of light quark masses are reported. 
We summarize these values in Table 1.

As shown in Table 1, there is not so a large discrepancy 
among these estimates as far as  $m_u$ and $m_d$ are concerned.
But, for the strange quark mass $m_s$, two different values, 
$m_s \simeq 175$ MeV and $m_s \simeq 200$ MeV have been reported. 
We use weighted averages as input values in our calculations.

\begin{center}
Table 1~. 
Light quark mass values at 1 GeV (in unit of MeV)
\end{center}%
\vspace{-5mm}
$$
\begin{array}{|c|ll|ll|ll|}\hline
 & \multicolumn{2}{c|} {m_u} & \multicolumn{2}{c|} 
{m_d} 
& \multicolumn{2}{c|} {m_s} \\
 \hline
\mbox{\rm Gasser and Leutwyler (1982)[1]}   & 5.1 & \pm 1.5 & 8.9 & \pm 2.6 & 175 & \pm 55 \\
 \hline
 \mbox{\rm Dominguez and Rafael (1987)[6]}  & 5.6 & \pm 1.1 & 9.9 & \pm 1.1 & 199 & \pm 33 \\
 \hline
\mbox{\rm Narison (1995)[7]} & 4 & \pm 1 & 10 & \pm 1 &  197 & \pm 29 \\
 \hline
\mbox{\rm Leutwyler (1996)[8]}  & 5.1 & \pm 0.9 & 9.3 & \pm 1.4 & 175 & \pm 25 \\
 \hline
\mbox{\rm Weighted averages }   & 4.90 & \pm 0.53 & 9.76 & \pm 0.63 & 187 & \pm 16 \\
 \hline
\end{array}
$$

\vglue.2in


\noindent {\bf 3. Review: pole masses of heavy quarks}

\vglue.1in

\noindent \underline{Charm and bottom quark masses}
\vglue.1in

Gasser and Leutwyler (1982) [1] have estimated charm and bottom 
quark masses $m_c$ and $m_b$ and
Tirard and Yudur\'{a}in (1994) [9] have re-estimated  $m_c$ and $m_b$ 
 precisely and rigorously.
On the other hand, from $\psi$- and $\Upsilon$-sum rules, 
Narison (1994) [10] has estimated the running quark masses
corresponding to the short-distance perturbative 
pole masses to two-loops and three loops.
In Table 2, we summarize their values.
We use weighted averages in Table 2 as input values in our calculations.

\begin{center}
Table 2~. 
Pole masses of charm and bottom quark
\end{center}%
\vspace{-5mm}
$$
\begin{array}{|c|ll|ll|}\hline
 & \multicolumn{2}{c|} {M_c^{pole}} & \multicolumn{2}{c|} 
{M_b^{pole}} \\
 \hline
\mbox{\rm Tirard and Yudur\'{a}in (1994)[9]}   & 1.570 & \pm 0.019 \mp 0.007 & 4.906 & _{-0.051}^{+0.069} \mp 0.004 _{-0.040}^{+0.011} \\
 \hline
\mbox{\rm Narison (1994)[10]}   & 1.64 & ^{+0.10}_{-0.07} \pm 0.03 & 4.87 & \pm 0.05\pm 0.02  \\
 \hline
\mbox{\rm Weighted averages}   & 1.59 & \pm 0.02 & 4.89 & \pm 0.05 \\
 \hline
\end{array}
$$

\vglue.1in

\noindent \underline{Top quark mass}
\newline

\noindent
\begin{minipage}[t]{\minitwocolumn}
The top quark mass values obtained by
by the CDF collaboration (1994) [5, 11] 
and the D0 collaboration [12] are summarized in Table 3.
 We use the values quoted by the particle data group (PDG96) [13] 
as the pole mass of the top quark.
\end{minipage}%
\hspace*{\columnsep}%
\begin{minipage}[t]{\minitwocolumn}
\begin{center}
Table 3~. 
Pole mass of top quark
\end{center}%
$$
\begin{array}{|c|lll|}\hline
 & \multicolumn{3}{c|} {M_t^{pole}} 
\\ \hline
\mbox{\rm CDF (1994)[5]}   & 174 & \pm 10 & ^{+13}_{-12}  \\
 \hline
\mbox{\rm CDF (1995)[11]}   & 176 & \pm 8 & \pm 10 \\
 \hline
\mbox{\rm D0 (1995)[12]}   & 199 & ^{+19}_{-21} & \pm 22 \\
 \hline
\mbox{\rm PDG (1996)[13]}   & 180 & \pm 12 & \\
 \hline
\end{array}
$$
\end{minipage}

\vspace{.2in}


\noindent {\bf 4. Estimation of $\alpha_s(\mu)$
and  $\Lambda_{\overline{\rm MS}}^{(n)}$}

\vglue.1in

Prior to estimates of the running quark masses $m_q(\mu)$, we must estimate
the values of $\alpha_s(\mu)$ and $\Lambda_{\overline{\rm MS}}^{(n)}$.
The effective QCD coupling $\alpha_s=g_s^2/4\pi$ is given by [14]

$$
\alpha_s (\mu)=\frac{4\pi}{\beta_0}\frac{1}{L}\left\{1-
\frac{2\beta_1}{\beta_0^2}\frac{\ln L}{L} 
+ \frac{4\beta_1^2}{\beta_0^4 L^2}
\left[ \left( \ln L - \frac{1}{2} \right)^2
+ \frac{\beta_2 \beta_0}{8\beta_1^2}  - \frac{5}{4} \right]
\right\} 
+O \left(\frac{\ln^2 L}{L^3} \right)
\ , 
\eqno(4.1)
$$
where
$$
\beta(\alpha_s)=-\frac{\beta_0}{2\pi}\alpha_s^2
-\frac{\beta_1}{4\pi^2} \alpha_s^3 
-\frac{\beta_2}{64\pi^3} \alpha_s^4
+O(\alpha_s^5) 
\ , 
\eqno(4.2)
$$
$$
\beta_0=11-\frac{2}{3}n_q \ , \ \ \ 
\beta_1=51-\frac{19}{3}n_q \ , \ \ \ 
\beta_2=2857-\frac{5033}{9}n_q +\frac{325}{27}n_q^2 
\ , 
\eqno(4.3)
$$
$$
L=\ln(\mu^2/\Lambda^2) \ .
\eqno(4.4)
$$ 

The value of
$\alpha_s(\mu)$ is not continuous at $n$th quark threshold $\mu_n$ at 
which the $n$th quark flavor channel is opened, because the coefficients
$\beta_0$, $\beta_1$ and $\beta_2$ 
depend on the effective quark flavor number $n_q$.

\noindent 
\begin{minipage}[tl]{5cm}
Therefore, 
we use the expression $\alpha_s^{(n)}(\mu)$ 
 with a different 
$\Lambda_{\overline{\rm MS}}^{(n)}$ for each energy scale range
$\mu_n\leq\mu<\mu_{n+1}$.

The values of $\Lambda_{\overline{\rm MS}}^{(n)}$ are evaluated 
by matching condition [4].
In Table 4, the values of $\Lambda_{\overline{\rm MS}}^{(n)}$
are summarized and 
the underlined values denote input values
$\Lambda_{\overline{\rm MS}}^{(5)}$.

We show the threshold behavior of $\alpha_s^{(n)} (\mu)$ in Fig.~1. 
We can see that $\alpha_s^{(n-1)} (\mu)$ in $\mu_{n-1}\leq\mu<\mu_n$ 
connects with $\alpha_s^{(n)} (\mu)$ in $\mu_n \leq \mu <\mu_{n+1}$ 
continuously.

\end{minipage}%
\hspace{\columnsep}%
\begin{minipage}[tr]{9cm}
{\epsfxsize=9.54cm \epsfysize=9.54cm \epsfbox{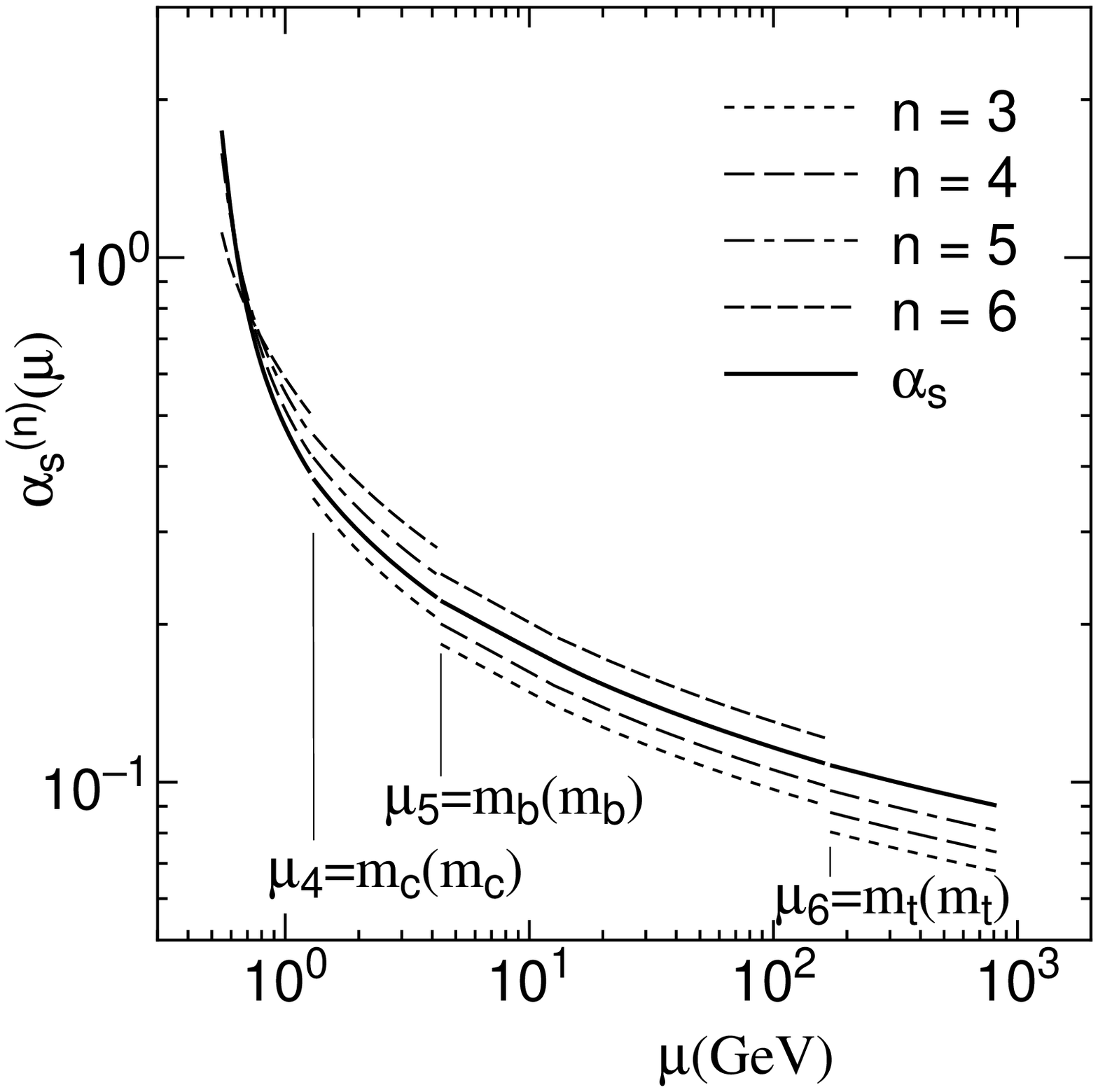}}
\vspace{-5mm}
\begin{quote}
Fig.~1. The threshold behavior of $\alpha_s^{(n)} (\mu)$ 
\end{quote}%
\end{minipage}%

\begin{center}
Table 4~. The values of $\Lambda_{\overline{\rm MS}}^{(n)}$ 
in unit of GeV.
\end{center}
\vspace{-5mm}
$$
\begin{array}{|c|c|c|c|c|}\hline
n & 3 & 4 & 5 & 6 
\\ \hline
\Lambda^{(n)}_{\overline{\rm MS}}  & 0.333^{+0.047}_{-0.042} & 
0.291^{+0.048}_{-0.041} &
\underline{0.209}^{+\underline{0.039}}_{-\underline{0.033}} &
 0.0882^{+0.0185}_{-0.0153} 
 \\ \hline
\end{array}
$$

\vglue.2in

\noindent {\bf 5.  Estimation of running quark masses $m_q(m_q)$}

\vglue.1in

{}From the pole mass values  $M^{pole}_q$, we estimate 
the running mass values  $m_q(\mu)$ at $\mu=M^{pole}_q$ 
for  $q=u,d,s$ by using the relation [15]
$$
\left. m_q(M_q^{pole})=M_q^{pole}\right/\left[1
+\frac{4}{3}\frac{\alpha_s(M_q^{pole})}{\pi}+
K \left(\frac{\alpha_s(M_q^{pole})}{\pi}\right)^2 
+O(\alpha_s^3)\right] \ , \eqno(5.1)
$$
The values of $K$, $M^{pole}_q$ and $m_q(\mu)$ at $M_q^{pole}$ 
for $q=c,b,t$ are summarized in Table~5.

\begin{center}
Table 5~. 
Parameter K , Pole masses $M_q^{pole}$ and Running mass $m_q(\mu)$ 
at $M_q^{pole}$ 
\end{center}%
\vspace{-5mm}
$$
\begin{array}{|c|c|c|c|}\hline
   &  K & M_q^{pole} \mbox{ (GeV) } & 
m_q (M_q^{pole}) \mbox{ (GeV) } \\ \hline
c &  14.47 & 1.59 & 1.213 \\ \hline
b & 12.94 & 4.89 & 4.248 \\ \hline
t & 10.98 & 180 & 170.1 \\ \hline
\end{array}
$$

The scale dependence of a running quark mass $m_q(\mu)$ is governed by the 
equation [2] 
$$
\mu\frac{d}{d\mu} m_q(\mu) = -\gamma(\alpha_s)m_q(\mu) \ \ , 
\eqno(5.2)
$$
where 
$$
\gamma(\alpha_s) = \gamma_0\frac{\alpha_s}{\pi} + 
\gamma_1\left(\frac{\alpha_s}{\pi}\right)^2 + 
\gamma_2\left(\frac{\alpha_s}{\pi}\right)^3 + 
O(\alpha_s^4) \ \ ,
\eqno(5.3)
$$
$$
\gamma_0 = 2 \ \ , \ \ \ \gamma_1 = \frac{101}{12} - \frac{5}{18}n_q \ \ , 
\gamma_2 = \frac{1}{32}\left[1249 - \left(\frac{2216}{27} + 
\frac{160}{3}\zeta(3) 
\right)n_q - \frac{140}{81}n_q^2\right] \ \ . 
\eqno(5.4)
$$
The running quark mass $m_q(\mu)$ is given by 
$$
m_q(\mu) = R(\alpha_s) \widehat{m}_q \ \ , 
\eqno(5.5)
$$
$$
R(\alpha_s) = \left(\frac{\beta_0}{2}\frac{\alpha_s}{\pi} \right)^
{2\gamma_0/\beta_0} \left\{1 + \left(2\frac{\gamma_1}{\beta_0} - 
\frac{\beta_1\gamma_0}{\beta_0^2} \right)\frac{\alpha_s}{\pi} \right. 
$$
$$ 
\left. + \frac{1}{2}\left[\left(2\frac{\gamma_1}{\beta_0} - 
\frac{\beta_1 \gamma_0}
{\beta_0^2} \right)^2 + 2\frac{\gamma_2}{\beta_0} - 
\frac{\beta_1\gamma_1}{\beta_0^2} - \frac{\beta_2\gamma_0}{16\beta_0^2} + 
\frac{\beta_1^2\gamma_0}{2\beta_0^3} \right] 
\left(\frac{\alpha_s}{\pi} \right)^2 + O(\alpha_s^3)\right\} \ \ , 
\eqno(5.6)
$$
\noindent 
\begin{minipage}[tl]{5cm}
where $\widehat{m}_q$ is the renormalization group invariant mass 
which is independent of $\ln(\mu^2/\Lambda^2)$, $\alpha_s$ is 
given by (4.1) and $\beta_i \ (i= 0, 1, 2)$ are defined by (4.3). 

By using  $\Lambda_{\overline{\rm MS}}^{(n)}$,
we can evaluate $R^{(n)}(\mu)$ for $ \mu < \mu_{n + 1}$, 
where $\mu_n$ is the $n$th quark flavor threshold and we take 
$\mu_n = m_{qn}(m_{qn})$. 
We show the threshold behavior of $R^{(n)} (\mu)$ in Fig.~2. 
As shown in Fig.~2, the behavior of $R(\mu)$ is discontinuous at 
$\mu=\mu_n \equiv m_{q_n}(m_{q_n})$.

\end{minipage}%
\hspace{\columnsep}%
\begin{minipage}[tr]{9cm}
{\epsfxsize=9.54cm \epsfysize=9.54cm \epsfbox{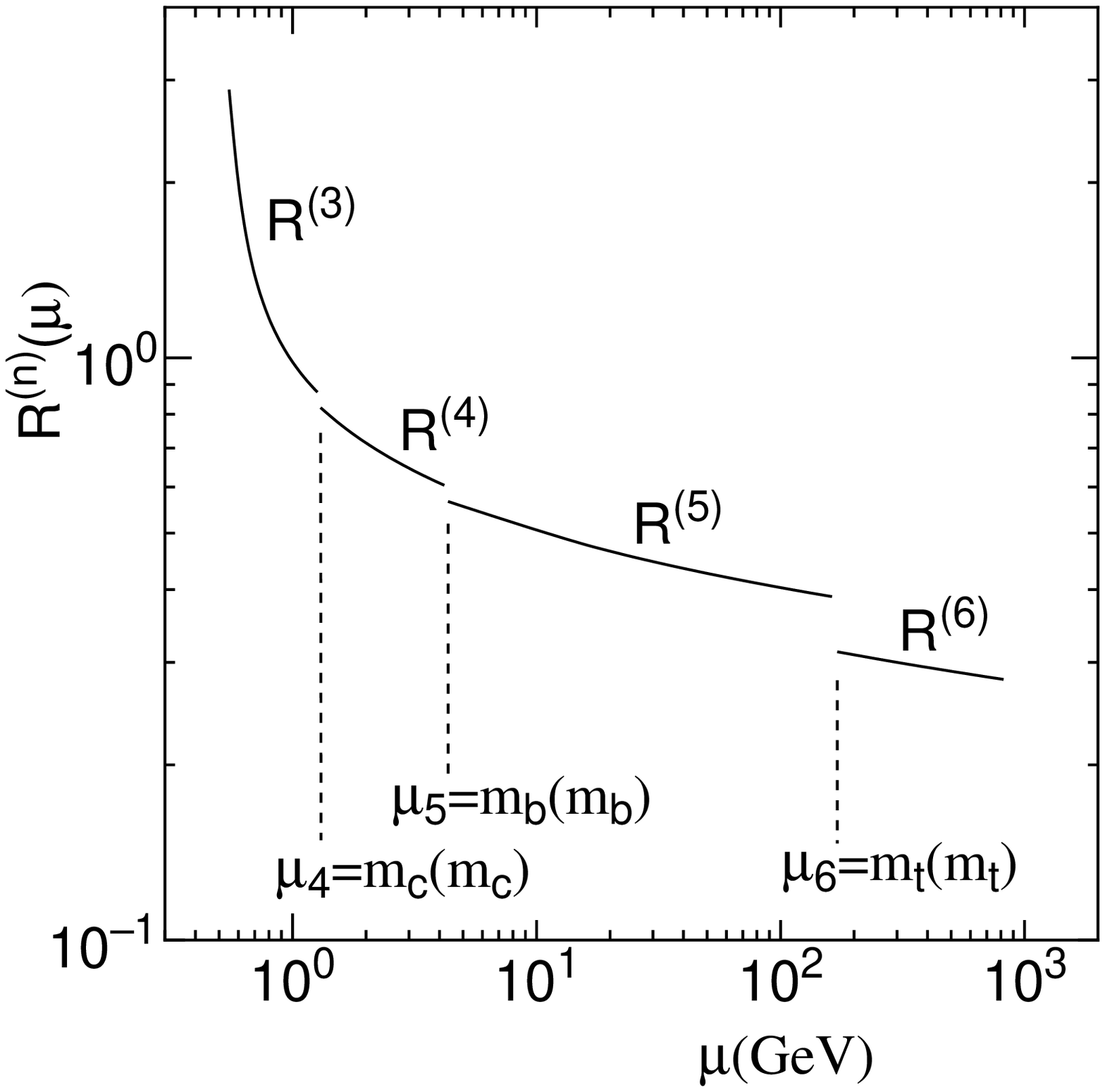}}
\vspace{-5mm}
\begin{quote}
Fig.~2. The threshold behaviour of $R^{(n)} (\mu)$ 
\end{quote}
\end{minipage}%


We can evaluate the values of $m_q(m_q)$ ($q=c,b,t$) 
by using the values of $M_q^{pole}$ 
and the relation 
$$
m_{qn}(\mu) = \left[R^{(n)}(\mu)/R^{(n)}(M_{qn}^{pole})\right] 
m_{qn}(M_{qn}^{pole}) \ \ \ (\mu < \mu_{n+1}) \ \ . 
\eqno(5.7)
$$

Similarly, we evaluate the light quark masses $m_q(m_q)$ 
 by using  the values $m_q(1 {\rm GeV})$ and the relation 

$$
m_q(\mu) = \left[R^{(3)}(\mu)/R^{(3)}(1{\rm GeV})\right] 
m_q(1 {\rm GeV}) \ \ \ (\mu < \mu_4) \ \ . 
\eqno(5.8)
$$


Running quark mass values $m_{qn}(\mu)$ at  $\mu \geq \mu_{n+1}$ 
cannot be evaluated by using $R(\mu)^{(n)}$ straightforwardly, 
because of the discontinuity of $R(\mu)$ at quark threshold 
$\mu=\mu_n \equiv m_{q_n}(m_{q_n})$.

The behavior of the $n$th quark mass $m_{qn}^{(N)} \ (n<N)$ 
at $\mu_N \leq \mu < \mu_{N + 1}$ is given by the matching condition 
[16]
$$
m_{qn}^{(N)} (\mu) =\left. m_{qn}^{(N-1)} (\mu)\right/ \left[1 + \frac{1}{12} 
\left(x_N^2 + \frac{5}{3}x_N + \frac{89}{36}\right)\left(\frac{\alpha_s^{(N)}
(\mu)}{\pi}\right)^2 \right] \ \ , 
\eqno(5.9)
$$
where 
$$
x_N = \ln \left[\left(m_{qN}^{(N)}(\mu)\right)^2/\mu^2\right] \ \ . 
\eqno(5.10)
$$

In Fig.~3, we illustrate the $\mu$-dependency of the light quark 
masses $m_q(\mu)$ ($q=u,d,s$) which take the matching condition 
(5.9) into account. 
We also illustrate the behavior of the heavy quark masses 
 $m_q(\mu)$ ($q=c,b,t$) in Fig.~4. 

\noindent
\begin{minipage}[tl]{\minitwocolumn}
{\epsfxsize=7.63cm \epsfysize=7.63cm \epsfbox{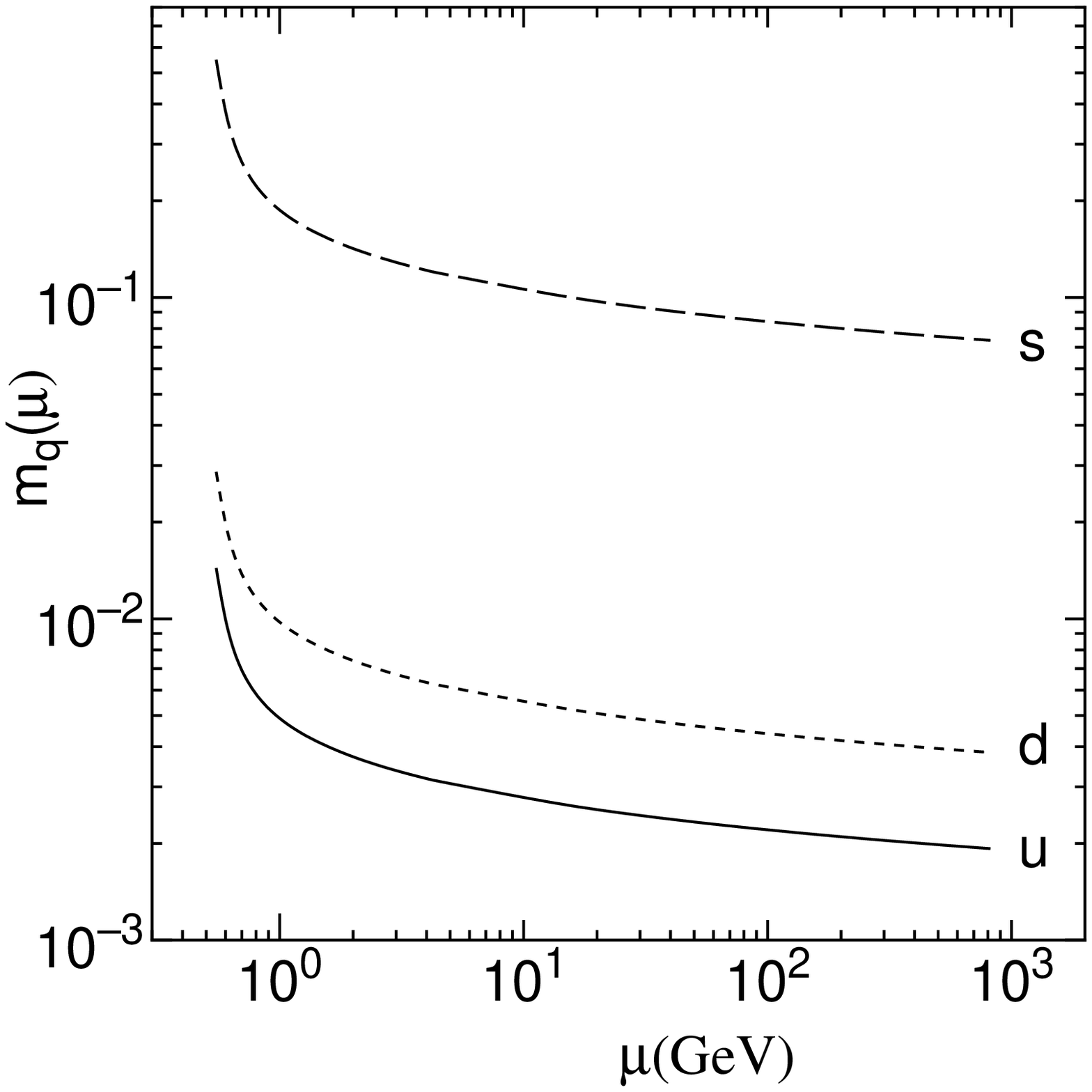}}
\vspace{-5mm}
\begin{quote}
Fig.~3. running masses of light quarks 
\end{quote}
\end{minipage}%
\hspace*{1mm}%
\begin{minipage}[tr]{\minitwocolumn}
{\epsfxsize=7.63cm \epsfysize=7.63cm \epsfbox{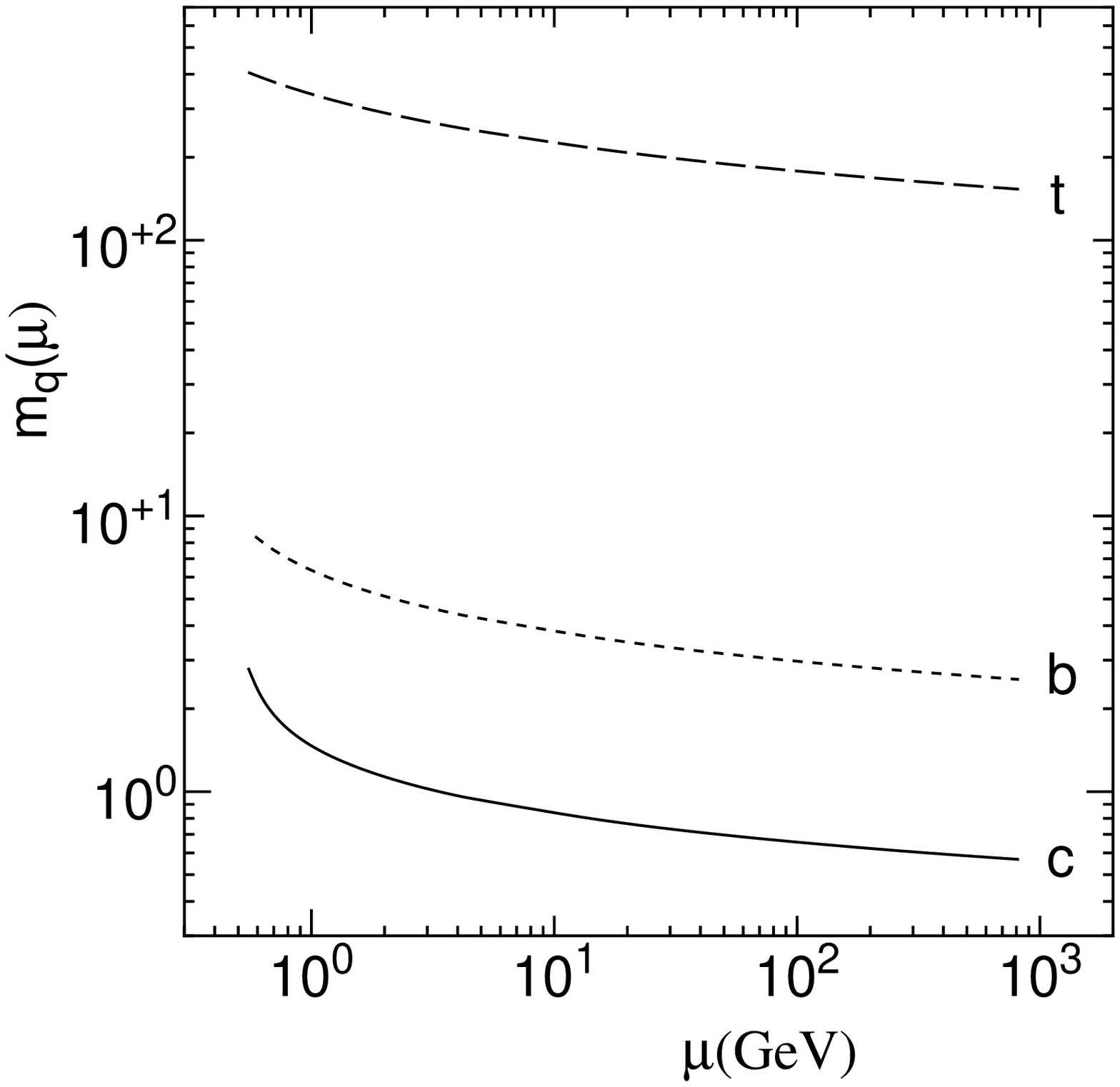}}
\vspace{-5mm}
\begin{quote}
Fig.~4. running masses of heavy quarks 
\end{quote}
\end{minipage}

\vspace{5mm}

The numerical results are summarized in Table 5, 
where input values $m_q$ (1 GeV) for $q = u, d, s$ 
and $m_q (M_q^{pole})$ for $q=c,b,t$ are used. 
The first and second errors come from $\pm\Delta m_q$ 
(or $\pm\Delta M_q^{pole}$) 
and $\pm\Delta\Lambda^{(5)}_{\overline{\rm MS}}$, respectively.
The values with asterisk should not be taken rigidly, 
because these values have been calculated in the region with 
a large $\alpha_s(\mu)$.

\begin{table}[htbp]
\begin{center}
Table 5~. 
Running quark mass values $m_q(\mu)$ at $\mu = m_q$
\end{center}%
$$
\begin{array}{|c|r|r|r|r|r|r|}\hline
q= & u \ \ \ \ & d \ \ \ \ & s \ \ \ \ & 
c \ \ \ \ & b \ \ \ \ & t \ \ \ \ \\[.1in] \hline
M_q^{pole} & * 0.501 & * 0.517 & * 0.681 & 1.59 & 4.89 & 180 \\ 
 & \pm 0.002 & \pm 0.002 & \pm 0.011 & \pm 0.02 & \pm 0.05 & \pm 12 \\ 
 & {}^{+0.058}_{-0.052} & {}^{+0.067}_{-0.060} & {}^{+0.065}_{-0.059} & 
 &  &  \\ \hline
m_q(M_q^{pole})  & * 0.0308 & * 0.0443 & * 0.275 & 1.213 & 4.248 & 170.1 \\ 
 & {}^{+0.0018}_{-0.0019} & {}^{+0.0016}_{-0.0017} & 
 {}^{+0.015}_{-0.016} & \pm 0.018 & \pm 0.046 & \pm 11.4 \\ 
 & {}^{+0.0002}_{-0.0005} & {}^{+0.0001}_{-0.0006} & {}^{-0.004}_{+0.001} & 
{}^{-0.040}_{+0.034} & {}^{-0.040}_{+0.036} & \mp0.3 \\ \hline
m_q(m_q)  & * 0.436 & * 0.448 & * 0.549 & 1.302 & 4.339 & 170.8 \\ 
 & {}^{+0.001}_{-0.002} & \pm 0.001 & \pm 0.007 & \pm 0.018 & \pm 0.046 & 
\pm 11.5 \\ 
 & {}^{+0.058}_{-0.052} & {}^{+0.059}_{-0.053} & {}^{+0.059}_{-0.052} & 
{}^{-0.020}_{+0.019} & {}^{-0.029}_{+0.027} & \mp0.2 \\ \hline
m_q({\rm 1 GeV}) & 0.00490 & 0.00976 & 0.187 
& 1.467 & 6.356 & 339 \\ 
 & \pm 0.00053 & \pm 0.00063 & \pm 0.016 & \pm 0.028 & \pm 0.080 & \pm 24 \\ 
   &  &  &   & {}^{+0.005}_{-0.002} & {}^{+0.214}_{-0.164} 
& {}^{+12}_{-11} \\ \hline
m_q(m_c) & 0.00421 & 0.00838 & 0.160 & 1.302 & 5.782 & 318 \\ 
m_c=1.302  & \pm 0.00045 & \pm 0.00054 & \pm 0.014 & \pm 0.018 
& \pm 0.047 & \pm 22 \\ 
  & {}^{-0.00011}_{+0.00007} & {}^{-0.00021}_{+0.00015} & {}^{-0.004}_{+0.003} 
& {}^{-0.020}_{+0.019} & {}^{+0.145}_{-0.112} & {}^{+10}_{-9} \\ \hline
m_q(m_b)   & 0.00312 & 0.00621 & 0.119 & 0.950 & 4.339 & 253 \\ 
m_b=4.339 & \pm 0.00034 & \pm 0.00040 & \pm 0.010 & \pm 0.016 
& \pm 0.046 & \pm 18 \\ 
 & {}^{-0.00020}_{+0.00016} & {}^{-0.00040}_{+0.00031} & 
{}^{-0.008}_{+0.006} & {}^{-0.052}_{+0.045} & {}^{-0.029}_{+0.027} & 
{}^{+4}_{-3} \\ \hline
m_q(m_W)   & 0.00224 & 0.00446 & 0.0855 & 0.668 & 3.029 & 182 \\ 
m_W=80.33 & \pm 0.00024 & \pm 0.00029 & \pm 0.0073 & \pm 0.013 
& \pm 0.038 & \pm 13 \\ 
 & {}^{-0.00017}_{+0.00014} & {}^{-0.00035}_{+0.00029} & 
{}^{-0.0066}_{+0.0055} & {}^{-0.047}_{+0.043} & {}^{-0.074}_{+0.069} & 
\pm 0.04 \\ \hline
m_q(m_Z)   & 0.00222 & 0.00442 & 0.0847 & 0.661 & 2.996 & 180 \\ 
m_Z=91.187 & \pm 0.00024 & \pm 0.00029 & \pm 0.0072 & \pm 0.012 
& \pm 0.038 & \pm 13 \\ 
  & {}^{-0.00017}_{+0.00014} & {}^{-0.00034}_{+0.00029} & 
{}^{-0.0066}_{+0.0055} & {}^{-0.047}_{+0.042} & {}^{-0.074}_{+0.069} & 
\pm 0.02 \\ \hline
m_q(m_t)   & 0.00212 & 0.00422 & 0.0809 & 0.630 & 2.847 & 170.8 \\ 
m_t=170.8 & \pm 0.00023 & \pm 0.00027 & \pm 0.0069 
& \pm 0.009 & {}^{+0.021}_{-0.020} & \pm 11.5 \\ 
 & {}^{-0.00017}_{+0.00014} & {}^{-0.00033}_{+0.00028} & 
{}^{-0.0063}_{+0.0053} & {}^{-0.045}_{+0.041} & {}^{-0.074}_{+0.069} & 
\mp 0.2 \\ \hline
m_q(\Lambda_W) & 0.00212 & 0.00422 & 0.0808 &  0.629 & 2.843 & 170.5  \\ 
\Lambda_W=174.1 & \pm 0.00023 & \pm 0.00027 & \pm 0.0069 & \pm 0.012 
& \pm 0.036 & \pm 12.3 \\ 
 & {}^{-0.00017}_{+0.00014} & {}^{-0.00033}_{+0.00028} & 
{}^{-0.0063}_{+0.0053} & {}^{-0.045}_{+0.041} & {}^{-0.075}_{+0.070} 
& \mp 0.3 \\ \hline
\end{array}
$$
\end{table}


\vglue .2in

\noindent {\bf 6.  
Reliability of the perturbative calculation below $\mu\sim 1$ GeV}

\vglue .1in

As we noted already, the values of the light quark masses $m_q(m_q)$ 
($q=u,d,s$) should not be taken rigidly, because the perturbative calculation
below $\mu\sim 1$ GeV seems to be not very reliable. 
Let us look at this more explicitly.

In order to see the reliability of the calculation of $\alpha_s(\mu)$ by 
using (4.1), in Fig.~5, we illustrate the values of the second 
and third terms in \{ \ \} of (4.1) separately. 
The values of the second and third terms exceed one at $\mu\simeq 0.5$ 
GeV and $\mu\simeq 0.6$ GeV, respectively.
Also, in Fig.~6, we illustrate the values of the second and third terms
in \{ \ \} of (5.6) separately. 
The values of the second and third terms exceed one at $\mu\simeq 0.6$ 
GeV and $\mu\simeq 0.7$ GeV, respectively.
These mean that the perturbative calculation is not reliable below 
$\mu\simeq 0.7$ GeV.
Therefore, the values with asterisk in Tables 5 should not be 
taken strictly.

\noindent
\begin{minipage}[tl]{\minitwocolumn}
{\epsfxsize=7.63cm \epsfysize=7.63cm \epsfbox{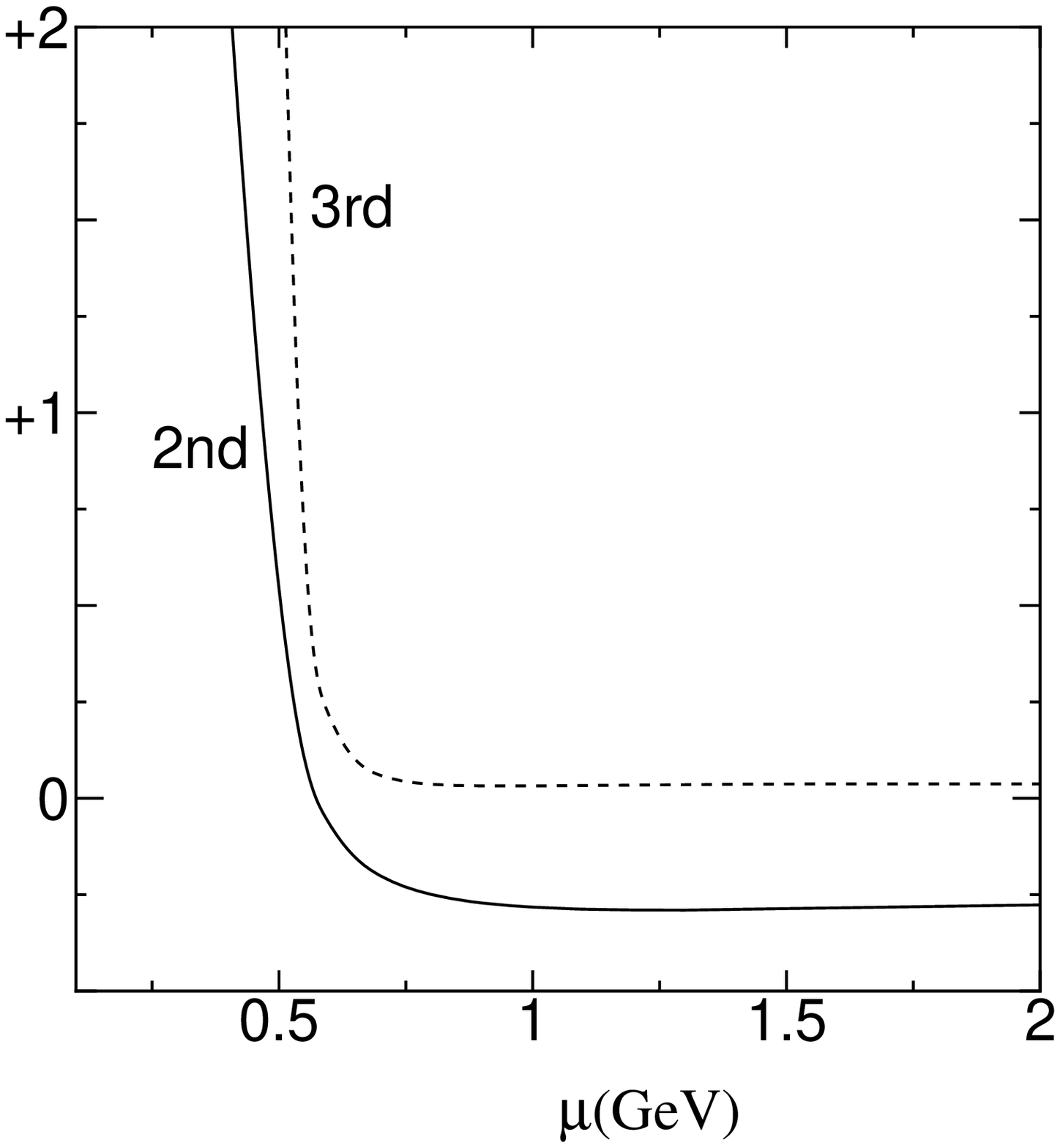}}
\vspace{-5mm}
\begin{quote}
Fig.~5. the values of the second and third term of (4.1)
\end{quote}
\end{minipage}%
\hspace*{1mm}%
\begin{minipage}[tr]{\minitwocolumn}
{\epsfxsize=7.63cm \epsfysize=7.63cm \epsfbox{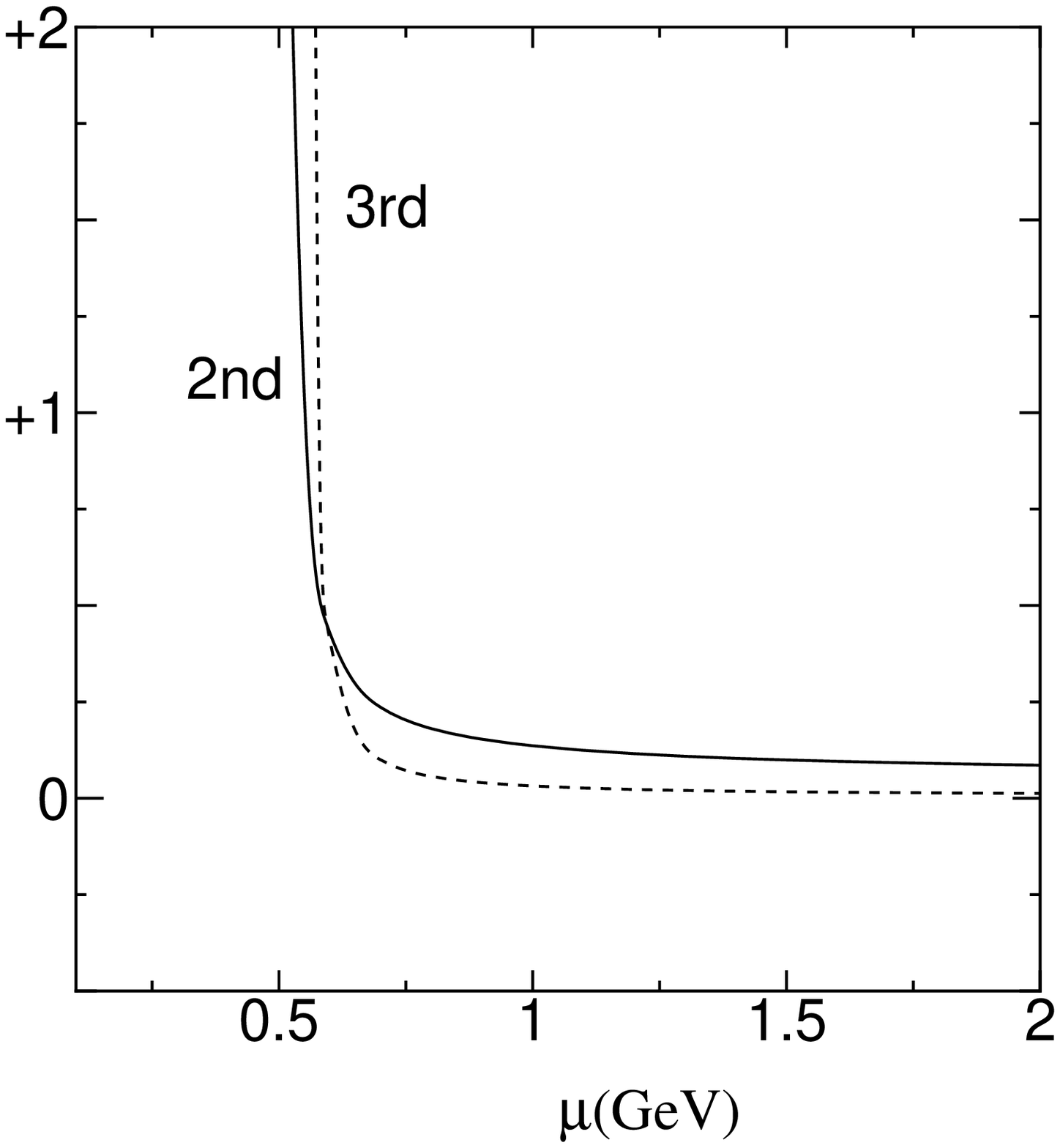}}
\vspace{-5mm}
\begin{quote}
Fig.~6. The values of the second and third term of (5.6) 
\end{quote}
\end{minipage}

\vglue.2in


\noindent {\bf 7. Evolution of Yukawa coupling constants}

\vglue.1in

By using the renormalization group equation,  
we estimate the Yukawa coupling constants 
in the standard model with a single Higgs boson.

The quark mass matrices 
$M_u$ and $M_d$ at $\mu=\Lambda_W$ are given by 
$$
M_a (\mu) = \frac{1}{\sqrt{2}} Y_a (\mu) v_a \ \ , 
\eqno(7.1)
$$
where $Y_a$ denotes a matrix of the Yukawa coupling constants $y^a_{ij}$ 
($a=u, d; i=1, 2, 3$), 
$(Y_a)_{ij}= y^a_{ij}$ and $v$ is the vacuum expectation value of
the Higgs boson.

The renormalization scale dependence of a matrix 
$
H_a = Y_a Y_a^\dagger 
$
 is given by [3]
$$
\frac{d}{dt} H_a = \left[ \frac{1}{16 \pi^2} \beta_a^{(1)} 
+ \frac{1}{(16\pi^2)^2} \beta_a^{(2)} \right] H_a 
+ H_a \left[\frac{1}{16\pi^2} \beta_a^{(1)\dagger} 
+ \frac{1}{(16\pi^2)^2} \beta_a^{(2) \dagger} \right] \ \ . 
\eqno(7.2)
$$
where $t$ is given by
$t = \ln (\mu/m_Z) $ 
and the one-loop and two-loop contributions 
$\beta_a^{(1)}$ and $\beta_a^{(2)}$ 
are written as 
$\beta_a^{(1)} = c_a^{(1)} {\bf 1} + \sum_b a_a^b H_b$ 
and
$\beta_a^{(2)} = c_a^{(2)} {\bf 1} + \sum_b b_a^b H_b + 
\sum_{b,c} b_a^{bc} H_bH_c $, 
respectively. The expressions of coefficients $c_a^{(i)}$, 
$ a_a^b$, etc. have been given in Ref.~[3].
For the input parameters, 
we use the quark masses $m_q(m_z)$ 
in Table~5 and 
the following parameters in the CKM matrix $V$ [13]:  
$$
|V_{us}| = 0.2205 \pm 0.0018  ,  \quad 
|V_{cb}| = 0.041 \pm 0.003  , \quad 
|V_{ub}/V_{cb}| = 0.08 \pm 0.02 \ . 
\eqno(7.3)
$$
For the gauge coupling constants, we use [17]
$$
\begin{array}{ll}
\alpha(m_Z)=(128.89 \pm 0.09)^{-1}  , \quad &
\sin^2\theta_W = 0.23165 \pm 0.000024  , \\    
\alpha_3(m_Z) = 0.118 \pm 0.003 \ . &
\end{array}
\eqno(7.4)
$$
The input value of Higgs boson $m_H$ is $\sqrt{2} \Lambda_W = 246.2$ GeV.

\noindent
\begin{minipage}[tr]{5cm}
If the input value $m_H$ is less than $2.2 \times 10^2$ GeV 
($2.3 \times 10^2$ GeV)
for two (one) loop evaluations, then 
the quartic coupling constant $\lambda$, 
of the Higgs boson self interaction, 
becomes negative at high energy.
On the other hand, if the input value $m_H$ is more than 
$2.6 \times 10^2$ GeV for both two and one loop calculations, 
the burst of $\lambda$ occurs at high energy.
In Fig.~7, we illustrate the $\mu$-dependency of the Yukawa coupling 
constants $y_q(\mu)$ ($q=u,d,s,c,b,t$) which take 
the renormalization equation into account. 
\end{minipage}
\hspace*{0.8\columnsep}%
\begin{minipage}[tl]{9cm}
{\epsfxsize=9.54cm \epsfysize=9.54cm \epsfbox{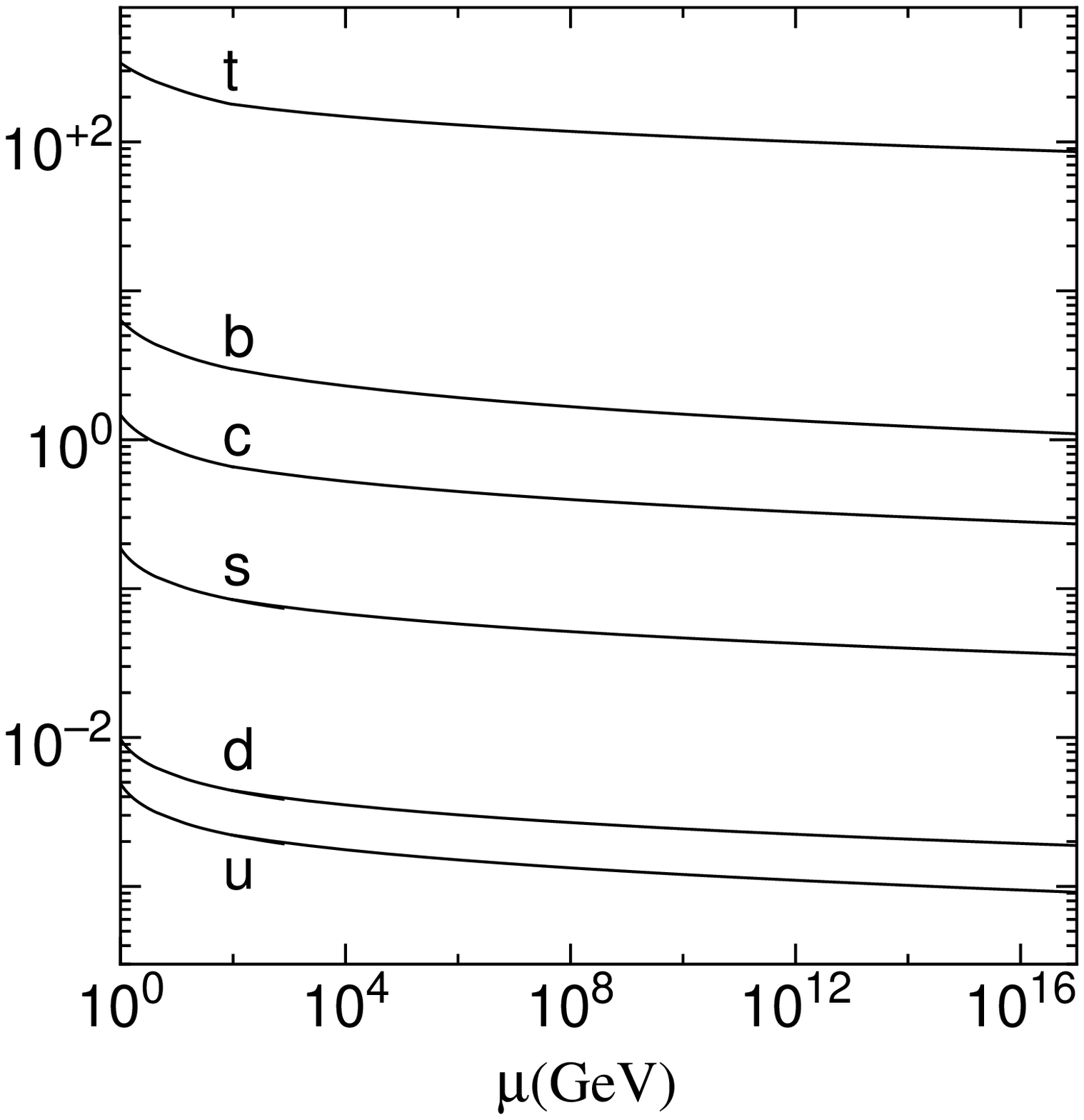}}
\vspace{-5mm}
\begin{quote}
Fig.~7. Evolution of the Yukawa coupling constants
\end{quote}
\end{minipage}%
\vspace{5mm}

\vglue.2in

\noindent {\bf 8. Summary}

\vglue.1in

We have evaluated the running quark mass values $m_q(\mu)$ 
at various energy scales below $\mu = \Lambda_W$ 
and the Yukawa coupling constants of the standard model
with a single Higgs boson at energy scales above  $\mu=\Lambda_W$. 
Although we have used the renormalization equation, 
the perturbative calculation
below $\mu\sim 0.7$ GeV is not adequately reliable
because the values of the second and third terms 
in the \{ \ \} of perturbative series (4.1) and (5.6) 
exceed one less than  $\mu\simeq 0.7$ 
GeV.

We discuss the grade of parameters 
fitted in  mass matrix models.  
At present, the ``confidence" grade of the ``observed" values 
of the running quark masses $m_q(\mu)$ and 
CKM matrix parameters are not equal at the same levels 
because these values are highly dependent on models 
or other experimental values (input values). 
Therefore, it is important for the model-building of 
quark mass matrix that we know the confidence levels of 
these values.  Our opinion based on the present work is 
summarized in the following Table:

\begin{center}
Table 6~. The reliability of the ``observed" values 
\end{center}%
\vspace{-5mm}
$$
\begin{array}{|ll|c|ccc|}\hline
  \multicolumn{2}{|c|} {\mbox{\rm grade}} & \mbox{\rm CKM matrix element}
& \multicolumn{3}{c|} {\mbox{\rm quark mass ratio}} \\
 \hline
 \mbox{(i)} & \mbox{(Reliable)} & |V_{us}| & & &   \\
 \hline
 \mbox{(ii)} & \mbox{(Almost reliable)} & |V_{cb}| & m_d/m_s & m_c/m_b &
  m_b/m_t  \\
 \hline
 \mbox{(iii)} & \mbox{(Somewhat variable)} & |V_{ub}| & m_c/m_t & m_u/m_c & 
 m_s/m_c  \\
 \hline
 \mbox{(iv)} & \mbox{(Variable)} &         & m_u/m_d & m_d/m_b &
  m_s/m_b  \\
 \hline
 \mbox{(v)} & \mbox{(Unreliable)}  & |V_{td}| &         &      
   &             \\
 \hline
\end{array}
$$
We have classified the CKM matrix elements
on the basis of the experimental and theoretical 
errors. 
In grading the quark mass ratios, 
we have considered ratios are reliable in the cases where 
(1) both two quarks are  heavy quarks or  light quarks
and (2) the mass difference between two quarks is small. 

Finally, we would like to point out that we should use 
the running mass values of $\mu=m_Z$ rather than $\mu=1$ GeV
for quark mass matrix phenomenology, together with the CKM matrix
parameters at $\mu=m_Z$. 

\vglue.3in

\begin{center}
{\large\bf Acknowledgments}\\[.1in]
\end{center}

The authors would like to express their sincere thanks to
 Prof.~M.~Tanimoto for his helpful discussions 
and Prof.~Z.~Hioki for informing the new values of electroweak 
parameters. 
This work was supported by the Grant-in-Aid for Scientific Research, 
Ministry of Education, Science and Culture, Japan (No.08640386).

\vglue.3in
\newcounter{0000}
\centerline{\bf References}
\begin{list}
{[~\arabic{0000}~]}{\usecounter{0000}
\labelwidth=1cm\labelsep=.4cm\setlength{\leftmargin=1.7cm}
{\rightmargin=.4cm}}
\item J.~Gasser and H.~Leutwyler, Phys.~Rep. {\bf 87}, 77 (1982).
\item O.~V.~Tarasov, Dubna preprint JINR P2-82-900, 1982 (unpublished). 
\item T.~P.Cheng, E.~Eichten and L.~F.~Li, Phys.~Rev. {\bf D9}, 
2259 (1974); 
M.~Machacek and M.~Voughn, Nucl.~Phys. {\bf B236}, 221 (1984).
\item W.~Bernreuther, Ann.~Phys. {\bf 151} , 127 (1983); 
I.~Hinchiliffe, p.77 
in Particle data group, R.~M.~Barnet {\it et al}., Phys.~Rev. {\bf D54} , 1 
(1996).
\item  CDF Collaboration, F.~Abe $et$ $al$., Phys.~Rev.~Lett. 
{\bf 73}, 225 (1994).
\item C.~A.~Dominguez and E.~de Rafael, Ann.~Phys. {\bf 174}, 372 (1987).
\item  S.~Narison, Phys.~Lett. {\bf B358} , 113 (1995).
\item H.~Leutwyler, Phys.~Lett. {\bf B378} , 313 (1996).
\item  S.~Titard and F.~J.~Yudur\'{a}in, Phys.~Rev. {\bf D49}, 6007 (1994).
\item S.~Narison, Phys.~Lett. {\bf B341} , 73 (1994). 
\item CDF collaboration, F.~Abe {\it at al}. , Phys.~Rev.~Lett. {\bf 74} , 
2626 (1995). 
\item D0 collaboration, S.~Abachi {\it et al}. , Phys.~Lett. {\bf 74} , 
2632 (1995).
\item Particle data group, R.~M.~Barnet {\it et al}., Phys.~Rev. {\bf D54} , 1 
(1996).
\item O.~V.~Tarasov, A.~A.~Vladimirov and A.~Yu.~Zharkov, Phys.~Lett. 
{\bf B93} , 429 (1980). 
\item  N.~Gray, D.~J.~Broadhurt, W.~Grafe and K.~Schilcher, Z.~Phys. 
{\bf C48}, 673 (1990).
\item W.~Bernreuther and W.~Wetzel, Nucl.~Phys. {\bf B197} , 228 (1982);
W.~Bernreuther, Ann.~Phys. {\bf 151} , 127 (1983);  
S.~A.~Larin, T.~van Ritbergen and J.~A.~M.~Vermaseren, Nucl.~Phys. 
{\bf B438} , 278 (1995). 
\item W.~Hollik, Invited talk at 11th Topical Workshop on 
Proton-Antiproton Collider Physics, Padua, Italy, May 26 -- June 1, 
1996, Univ.~Karlsruhe preprint KA-TP-19-1996 (1996). 
See, also, Z.~Hioki, Act.~Phys.~Polonica {\bf B27}, 1569 (1996).

\end{list}

\end{document}